\documentclass{IEEEtran}
\usepackage{cite}
\usepackage{amsmath,amssymb,amsfonts}
\usepackage{algorithmic}
\usepackage{graphicx}
\usepackage{textcomp}
\def\BibTeX{{\rm B\kern-.05em{\sc i\kern-.025em b}\kern-.08em
		T\kern-.1667em\lower.7ex\hbox{E}\kern-.125emX}}

\usepackage{color}
\usepackage{bm}
\usepackage{algorithm}

\usepackage{hhline}

\ifCLASSOPTIONcompsoc
\usepackage[caption=false, font=normalsize, labelfont=sf, textfont=sf]{subfig}
\else
\usepackage[caption=false, font=footnotesize]{subfig}
\fi

\usepackage{soul}

\usepackage{babel}
\usepackage{threeparttable}

\begin{document}
	\title{{\fontsize{24}{26}\selectfont{Communication\rule{29.9pc}{0.5pt}}}\break\fontsize{16}{18}\selectfont
		Near-Field Millimeter-Wave Imaging via Circular-Arc MIMO Arrays}
	
	\author{Shiyong Li,  
		Shuoguang Wang,
		Guoqiang Zhao,
		and Houjun Sun
		\thanks{
			The work was supported by the National Natural Science Foundation of China under Grant 61771049.  }
		
		\thanks{The authors are with the Beijing Key Laboratory of Millimeter Wave and Terahertz Technology, Beijing Institute of Technology, Beijing 100081, China. (e-mail: lisy\_98@bit.edu.cn).}
		
	}
	
	\maketitle

	\begin{abstract}
		Millimeter-wave (MMW) imaging has a wide prospect in application of concealed weapons detection. 
		We propose a circular-arc multiple-input multiple-output (MIMO) array scheme with uniformly spaced transmit and receive antennas along the horizontal-arc direction, while scanning along the vertical direction.   
		The antenna beams of the circular-arc MIMO array can provide more uniform coverage of the imaging scene than those of the linear or planar MIMO arrays. 
		Further, a  near-field three-dimensional (3-D) imaging algorithm, based on the spatial frequency domain processing, is presented 
		with analysis of sampling criteria and resolutions. 
		Numerical simulations, as well as comparisons with the back-projection (BP) algorithm, are provided to show the efficacy of the proposed approach.

	\end{abstract}
	
	
	\IEEEpeerreviewmaketitle
	
	\section{Introduction}
	Millimeter-wave (MMW) imaging can provide high resolutions of the target under test. Thus, it is of interest in a wide applications, such as remote sensing \cite{remote_sensing}, biomedical diagnosis \cite{biomedical}, and indoor target tracking \cite{Amin_2010book}. 
	In recent years, there is an increasing demand for MMW to detect  concealed weapons and contraband carried by personnel \cite{sheen,sheen1,ahmed1,zhuge2}, due to the fact that MMW can penetrate regular clothing to form an image of a person as well as the concealed objects with no health hazard at moderate power levels. 
	
	Usually, the antenna apertures are formed by 1-D monostatic arrays accompanied by mechanical scanning. The monostatic arrays need to satisfy the Nyquist sampling criterion, which leads to high number of antennas and switches.
	State-of-the-art MMW imaging systems employ multiple-input multiple-output (MIMO) arrays to reduce the number of elements \cite{subsampled_array,ahmed1,zhuge2}. Another merit of using MIMO lies in the fact that it can reduce the reconstructed artifacts induced by multipath reflections, which are prevalent in
	the monostatic array systems \cite{sheen1,ghosts_insight,wavenumber1}.
	
	Imaging systems combining  1-D linear MIMO arrays and synthetic aperture radar (SAR) were discussed in \cite{zhuge1,1dmimo_2, 1dmimo, 1d_mimo_cylindrical,guangyou} for concealed weapons detection.
	The two-dimensional (2-D) MIMO arrays with different topologies were studied in \cite{qps,zhuge_ap,zhuge2,2dmimo,tankai2,nufft_mimo2d} for near-field imaging. In so doing, the mechanical scanning is eliminated, which acheives real-time data acquisition. 
	However, the cost and complexity of 2-D MIMO array  systems are much higher than those constructed by  a 1-D array with mechanical scanning perpendicular to the array dimension.

	The 1-D ultrawideband (UWB) MIMO array was designed according to the effective aperture approach in \cite{zhuge1} associated with a time-domain imaging algorithm. 
	In \cite{1dmimo_2}, the 1-D linear MIMO arrays were optimized for short-range applications also on the basis of effective aperture approach. 
	In \cite{1dmimo}, imaging algorithms based on the modified Kirchhoff method were derived for linear MIMO arrays with receivers evenly distributed. 
	A spatial frequency domain imaging algorithm was developed in \cite{1d_mimo_cylindrical} for a linear MIMO array system associated with cylindrical scanning. 
	The extended phase shift migration algorithm was studied in \cite{guangyou} for a MIMO-SAR system working at Terahertz band.
	
	
	The aforementioned MIMO arrays are all configured with antennas placed on a straight line or a plane. 
	Usually, two separate dense transmit subarrays were set at both ends of the undersampled receive array for the 1-D MIMO array design. 
	This scheme can hardly provide an even illumination of a large imaging area along the array direction. 
	A single-frequency MIMO-arc array based azimuth imaging method was presented in \cite{mimo_arc_imaging}, based on the geometry transformation from the arc array to the equivalent linear array \cite{circular_linear}. 
	
	In this communication, we propose a circular-arc MIMO array based three-dimensional (3-D) imaging scheme. 
	The transmit and receive antennas are evenly placed along the horizontal-arc direction. And the array is mechanically scanned along its perpendicular direction to obtain a cylindrical aperture, as illustrated in Fig. \ref{circular_mimo}. 
	The antenna beams of the circular-arc MIMO array cover the imaging area more evenly 
	than those of the linear or planar MIMO arrays. 
	Furthermore, we present a near-field 3-D imaging algorithm based on the spatial frequency domain processing, to fully utilize the fast Fourier transforms (FFTs), for the circular-arc MIMO array system.
	The key to obtain the imaging algorithm is to solve the convolution of the Green's functions in the Fourier domain, which is, however, hard to be resolved. 
	We solve it through using an approximation of the radial distances between antennas and target pixels in the cylindrical coordinates. 
	Then, the spatial frequency domain imaging algorithm can be acquired based on the decomposition of the cylindrical waves into superposition of plane waves.

	The rest of the communication is organized as follows: In the next section, we formulate the imaging algorithm based on solving the convolution of the Green's functions in the spatial frequency domain. The sampling criteria and resolutions are  outlined. Numerical results are shown in Section III. And concluding remarks follow at the end.

	\section{Circular-Arc MIMO Array Based Imaging}
	
	\subsection{Imaging Algorithm}
	
	\begin{figure}[!t]
		\centering
		
		\includegraphics[width=2.0in]{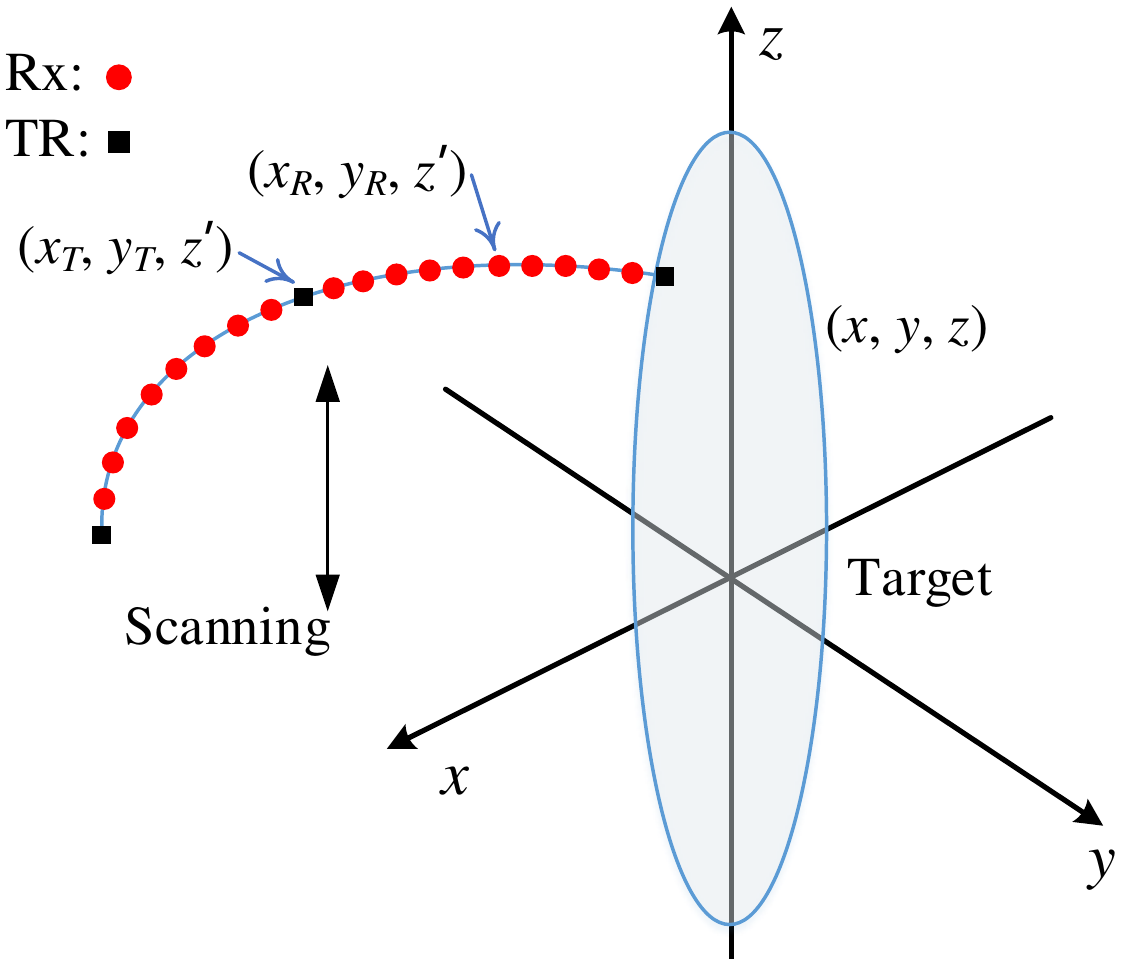}
		
		\caption{Topology of the circular-arc MIMO array based MMW imaging.}
		\label{circular_mimo}
	\end{figure}
	

	The circular-arc MIMO array based imaging geometry is shown in Fig. \ref{circular_mimo}, where the transmit and receive antennas are uniformly spaced along a circular arc, meanwhile, scanning along the vertical direction. 
	The demodulated scattered waves are given by, 
	\begin{equation}\label{scat_wave1}
		s(k,\theta_T,\theta_R,z')\!=\!\iiint \!\! g(x,y,z)e^{-\mathrm{j}k(R_T+R_R)}\mathrm{d}x\mathrm{d}y\mathrm{d}z, 
	\end{equation}
	where $k=\frac{2\pi f}{c}$ denotes the wavenumber, $f$ is the working frequency, $c$ is the speed of light, $g(x,y,z)$ represents the scattering coefficient of the target located at $(x,y,z)$, $R_T$ and $R_R$ are the distances from the transmit antenna to the target and from the target to the receive antenna, respectively, which can be expressed as,
	\begin{align}\nonumber
		R_T &=\sqrt{\rho_T^2+(z-z')^2},  \\ 
		R_R &=\sqrt{\rho_R^2+(z-z')^2},   \nonumber
	\end{align}
	where 
	\begin{equation}\label{rhoT}
		\rho_T=\sqrt{(x-R_0\sin\theta_T)^2+(y+R_0\cos\theta_T)^2},
	\end{equation}
	\begin{equation}\label{rhoR}
		\rho_R=\sqrt{(x-R_0\sin\theta_R)^2+(y+R_0\cos\theta_R)^2}.
	\end{equation}
	The transmit and receive positions are denoted by  $(R_0,\theta_T,z')$ and $(R_0,\theta_R,z')$, respectively, in the cylindrical coordinates, where $R_0$ represents the radius of the circular array.
	
	Next, we derive a spatial frequency domain algorithm from this imaging scheme. 
	Performing the Fourier transform on both sides of \eqref{scat_wave1} with respect to $z'$ and using the property of convolution in the Fourier domain, we have
	\begin{align}\label{scat_wave_kz0}
		&s(k,\theta_T,\theta_R,k_{z'})=\iiint g(x,y,z)\cdot \\ \nonumber
		&\mathcal{F}_{z'}[e^{-\mathrm{j}kR_T}]\circledast_{k_{z'}} \mathcal{F}_{z'}[e^{-\mathrm{j}kR_R}]\mathrm{d}x\mathrm{d}y\mathrm{d}z. 
	\end{align}
	The exponential terms $e^{-\mathrm{j}kR_T}$ and $e^{-\mathrm{j}kR_R}$  are referred to as the free space Green's functions, whose Fourier transforms with respect to $z'$ can be expressed as \cite{soumekh},
	\begin{equation} \label{grT}
		\mathcal{F}_{z'}[e^{-\mathrm{j}kR_T}]= e^{-\mathrm{j}\sqrt{k^2-k^2_{z'}}\rho_T}e^{-\mathrm{j}k_{z'}z},
	\end{equation}
	\begin{equation} \label{grR}
		\mathcal{F}_{z'}[e^{-\mathrm{j}kR_R}]= e^{-\mathrm{j}\sqrt{k^2-k^2_{z'}}\rho_R}e^{-\mathrm{j}k_{z'}z}.
	\end{equation}
	Substituting \eqref{grT} and \eqref{grR} in \eqref{scat_wave_kz0}, we obtain,
	\begin{align}\label{scat_wave_kz1}
		&s(k,\theta_T,\theta_R,k_{z'})=\iiint g(x,y,z)\cdot \\ \nonumber
		&[e^{-\mathrm{j}\sqrt{k^2-k^2_{z'}}\rho_T}e^{-\mathrm{j}k_{z'}z}\circledast_{k_{z'}} e^{-\mathrm{j}\sqrt{k^2-k^2_{z'}}\rho_R}e^{-\mathrm{j}k_{z'}z}]\mathrm{d}x\mathrm{d}y\mathrm{d}z. 
	\end{align}
	According to the  form of convolution, \eqref{scat_wave_kz1} is rewritten as,
	\begin{align}\label{scat_wave_kz2}
		&s(k,\theta_T,\theta_R,k_{z'})=\iiint g(x,y,z)e^{-\mathrm{j}k_{z'}z}\cdot \\ \nonumber
		&[e^{-\mathrm{j}\sqrt{k^2-k^2_{z'}}\rho_T}\circledast_{k_{z'}} e^{-\mathrm{j}\sqrt{k^2-k^2_{z'}}\rho_R}]\mathrm{d}x\mathrm{d}y\mathrm{d}z. 
	\end{align}
	
	However, it is hard to find the analytical solution of the convolution  in the square brackets. 
	In order to solve it, 
	we approximate $\rho_T\approx\rho_R$ (denoted by $\rho_0$) temporarily, according to the geometry information between the imaging scene (mainly located around the axis of a cylindrical surface constructed by the mechanical scanned array) and the circular-arc MIMO array. 
	Then, the convolution in the square brackets in \eqref{scat_wave_kz2} can be expressed as,
	\begin{align}\label{conv_krho}
		&e^{-j\sqrt{k^2-k^2_{z'}}\rho_T}\circledast_{k_{z'}} e^{-j \sqrt{k^2-k^2_{z'}}\rho_R} \\ \nonumber
		&\approx e^{-j\sqrt{k^2-k^2_{z'}}\rho_0}\circledast_{k_{z'}} e^{-j\sqrt{k^2-k^2_{z'}}\rho_0} \\ \nonumber
		&=\int e^{-j\sqrt{k^2-\zeta^2}\rho_0}e^{-j\sqrt{k^2-(k_{z'}-\zeta)^2}\rho_0}\mathrm{d}\zeta.
	\end{align}
	This integral can be calculated using the principle of stationary phase (POSP) \cite{cumming}. Assuming
	\begin{align}\label{phase}
		\vartheta(\zeta) = -\sqrt{k^2-\zeta^2}\rho_0-\sqrt{k^2-(k_{z'}-\zeta)^2}\rho_0,
	\end{align}
	then, using $\mathrm{d}\vartheta/\mathrm{d}\zeta = 0$, we obtain $\zeta = k_{z'}/2$. Thus, the convolution in \eqref{conv_krho} is given by,
	\begin{align}\label{conv_krho_1}
		&e^{-j\sqrt{k^2-k^2_{z'}}\rho_T}\circledast_{k_{z'}} e^{-j \sqrt{k^2-k^2_{z'}}\rho_R}\!\approx\! e^{j\vartheta(\zeta=k_{z'}/2)}e^{-j\pi/4}\\ \nonumber 
		&\approx e^{-j\sqrt{k^2-\frac{k_{z'}^2}{4}}\rho_T}e^{-j\sqrt{k^2-\frac{k_{z'}^2}{4}}\rho_R}e^{-j\pi/4}.
	\end{align}
	Note that we replace $\rho_0$ by the original $\rho_T$ and $\rho_R$ after the convolution. Here, we omit the envelope and the constant terms. 
	
\begin{figure}[!t]
		\centering
		
		\includegraphics[width=1.4in]{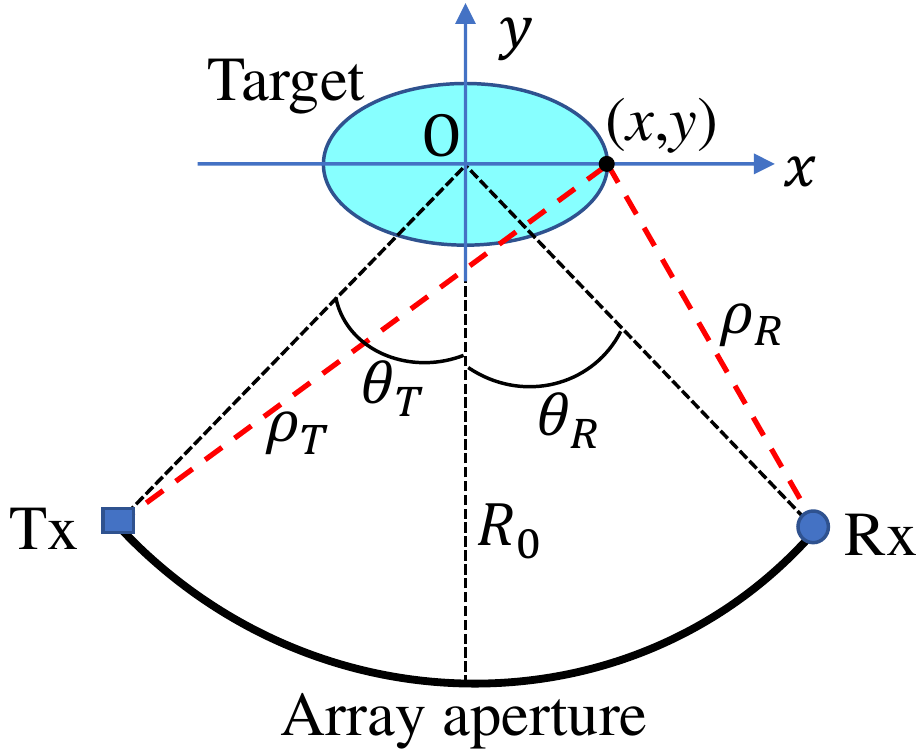}
		
		\caption{Geometry of a simple circular-arc MIMO array and target.}
		\label{illustration_geometry}
	\end{figure}

\begin{figure}[!t]
		\centering
		\subfloat[]{\label{a}
			\includegraphics[width=2.4in]{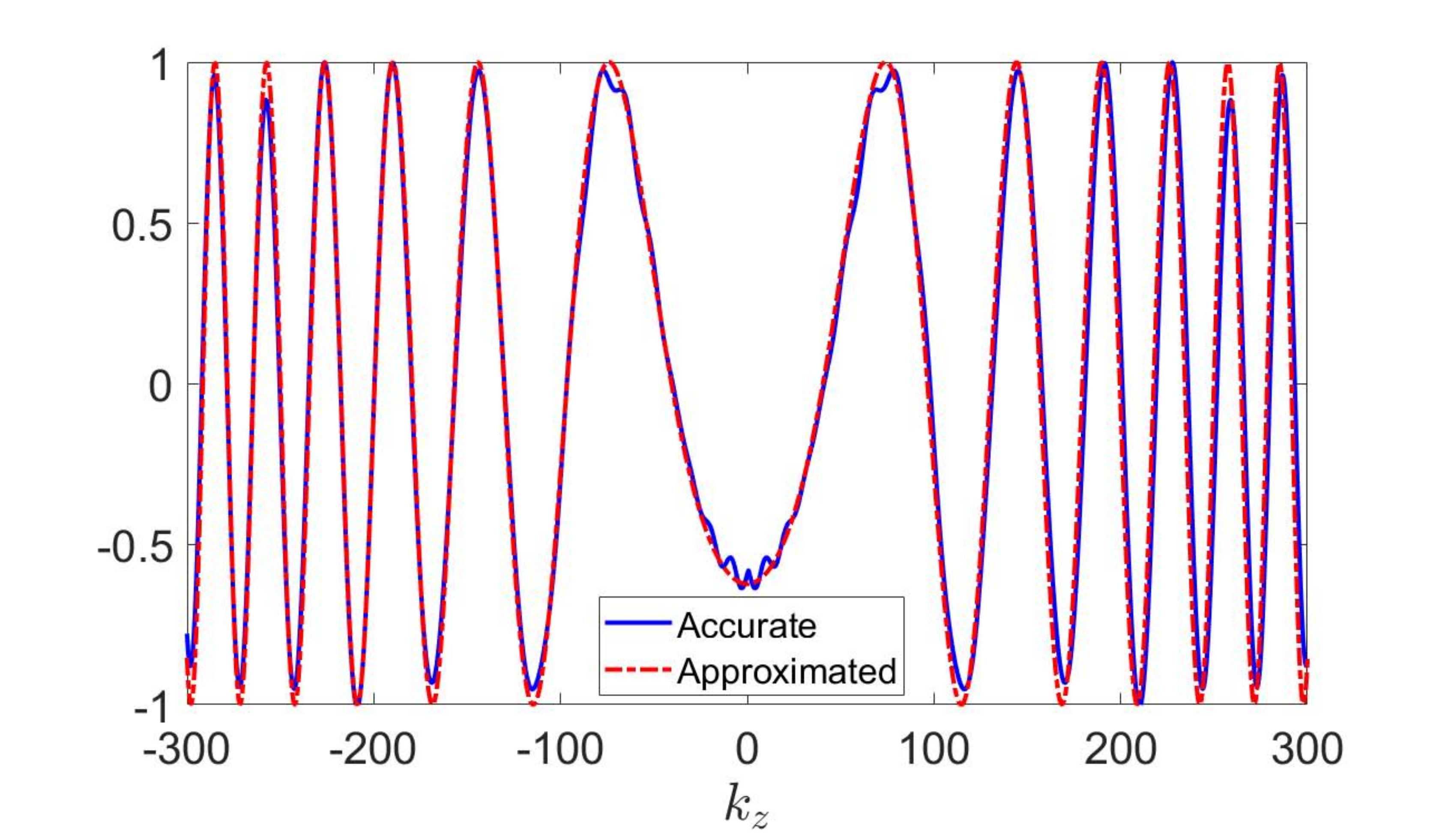}}
		\hfill
		\subfloat[]{\label{b}
			\includegraphics[width=2.4in]{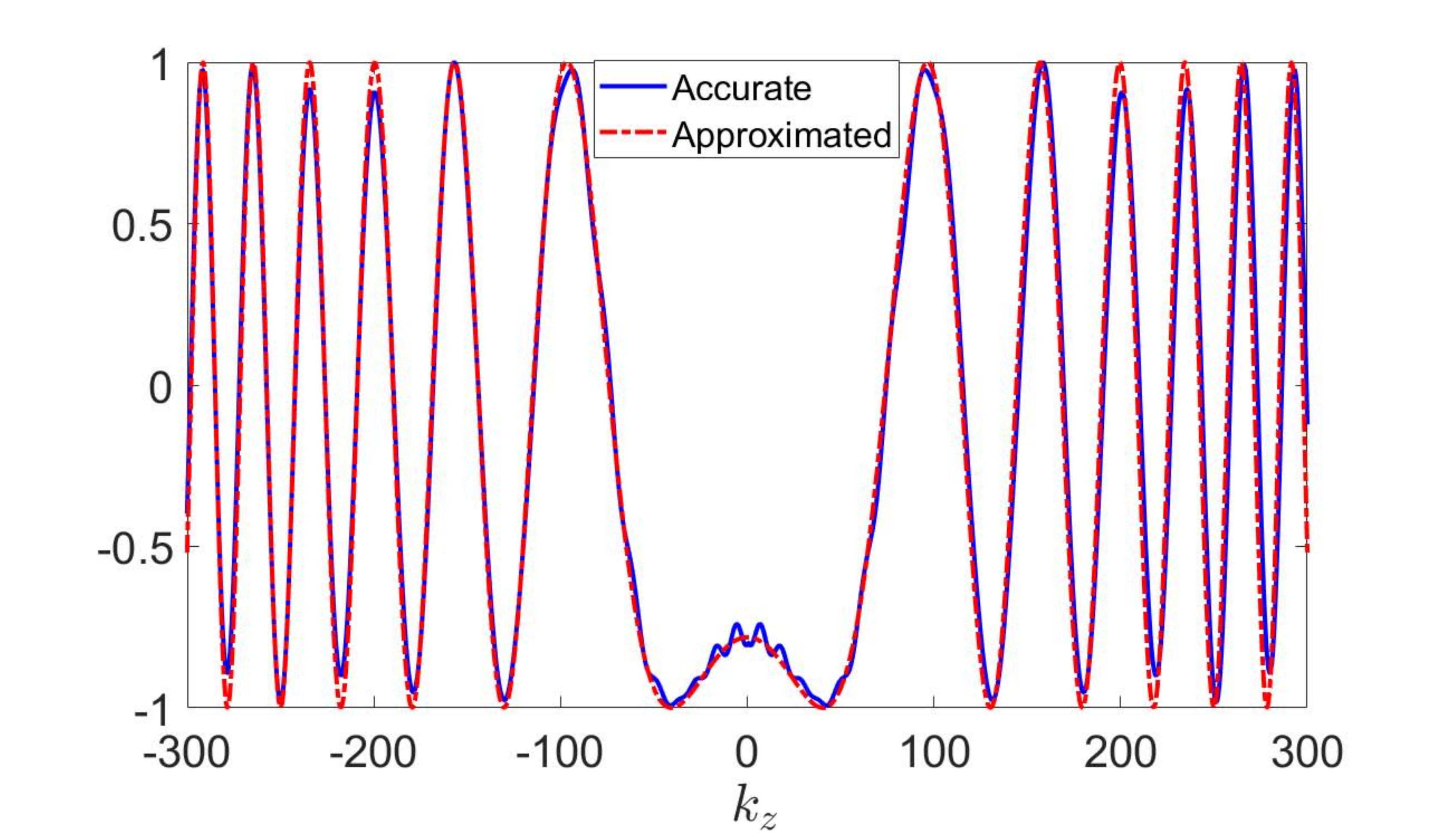}}
		\hfill
		\\	
		\caption{Comparison between the true convolution and the approximated one, (a) the real part, and (b) the imaginary part.}
		\label{conv_error}
	\end{figure}

	To evaluate the accuracy of \eqref{conv_krho_1}, we show a numerical comparison between $e^{-j\sqrt{k^2-k^2_{z'}}\rho_T}\circledast_{k_{z'}} e^{-j \sqrt{k^2-k^2_{z'}}\rho_R}$, which is calculated by the MATLAB function `conv' as the accurate values, and $e^{-j\sqrt{k^2-{k_{z'}^2}/{4}}\rho_T}e^{-j\sqrt{k^2-{k_{z'}^2}/{4}}\rho_R}e^{-j\pi/4}$, under a configuration illustrated in Fig. \ref{illustration_geometry} with parameters shown in Table \ref{tab1}.
	The results are shown in Fig. \ref{conv_error}. 
	Clearly, the errors induced by the approximation $\rho_T\approx\rho_R$    are really small in the process of solving the convolution. 
	
	\begin{table}
		\centering
		\caption{Parameters for Evaluating the Convolution}
		\label{tab3}
		\setlength{\tabcolsep}{3pt}
		\begin{threeparttable}
			\begin{tabular}{p{150pt}  p{80pt}}
				\hline\hline
				Parameters& Values \\[0.5ex]
				\hline
				Radius of the circular array $(R_0)$&
				1.0 m\\[0.5ex]
				working frequency& 
				30 GHz \\[0.5ex]
				Angle of transmit antenna $\theta_T$&
				$-20^{\circ}$ \\[0.5ex]
				Angle of receive antenna $\theta_R$&
				$20^{\circ}$ \\[0.5ex]
				Scanning step along $z$ direction $\Delta z$&
				1 cm \\[0.5ex]
				Position of the considered target pixel $(x,y)$&
				$(0.25,0)$ (Unit: meter) \\[0.5ex]
				
				\hline
			\end{tabular}
			\label{tab1}
		\end{threeparttable}
	\end{table}
	
	Substituting \eqref{conv_krho_1} in \eqref{scat_wave_kz2} yields, 
	\begin{align}\label{scat_wave_kz3}
		&s(k,\theta_T,\theta_R,k_{z'})=\iiint g(x,y,z)e^{-{j}k_{z'}z}\cdot \\ \nonumber
		&e^{-jk_\rho\rho_T}e^{-jk_\rho\rho_R}\mathrm{d}x\mathrm{d}y\mathrm{d}z,
	\end{align}
	where 
	\begin{equation} \label{dispersion}
		k_\rho = \sqrt{k^2-k_{z'}^2/4},
	\end{equation}
	and $e^{-j\pi/4}$ is omitted. 
	Based on \eqref{rhoT} and \eqref{rhoR}, we decompose the cylindrical waves  into a superposition of plane waves,
	\begin{equation} \label{cylindrical2planeT}
		e^{-\mathrm{j}k_{\rho} \rho_T}=\int e^{-\mathrm{j}k_{x_T}(x-x_T)}e^{-\mathrm{j}k_{y_T}(y-y_T)}\mathrm{d}k_{x_T},
	\end{equation}
	\begin{equation} \label{cylindrical2planeR}
		e^{-\mathrm{j}k_{\rho} \rho_R}=\int e^{-\mathrm{j}k_{x_R}(x-x_R)}e^{-\mathrm{j}k_{y_R}(y-y_R)}\mathrm{d}k_{x_R},
	\end{equation}
	where $x_{T/R}=R_0\sin\theta_{T/R}$, $y_{T/R}=-R_0\cos\theta_{T/R}$, and $k_{x_{T/R}}^2+k_{y_{T/R}}^2=k_{\rho_{T/R}}^2$ with $k_{\rho_{T/R}}=k_{\rho}$. Here, the subscript `$T/R$'  denotes the transmit or  receive parts for conciseness. 
	
\begin{figure}[!t]
		\centering
		\includegraphics[width=2.3in]{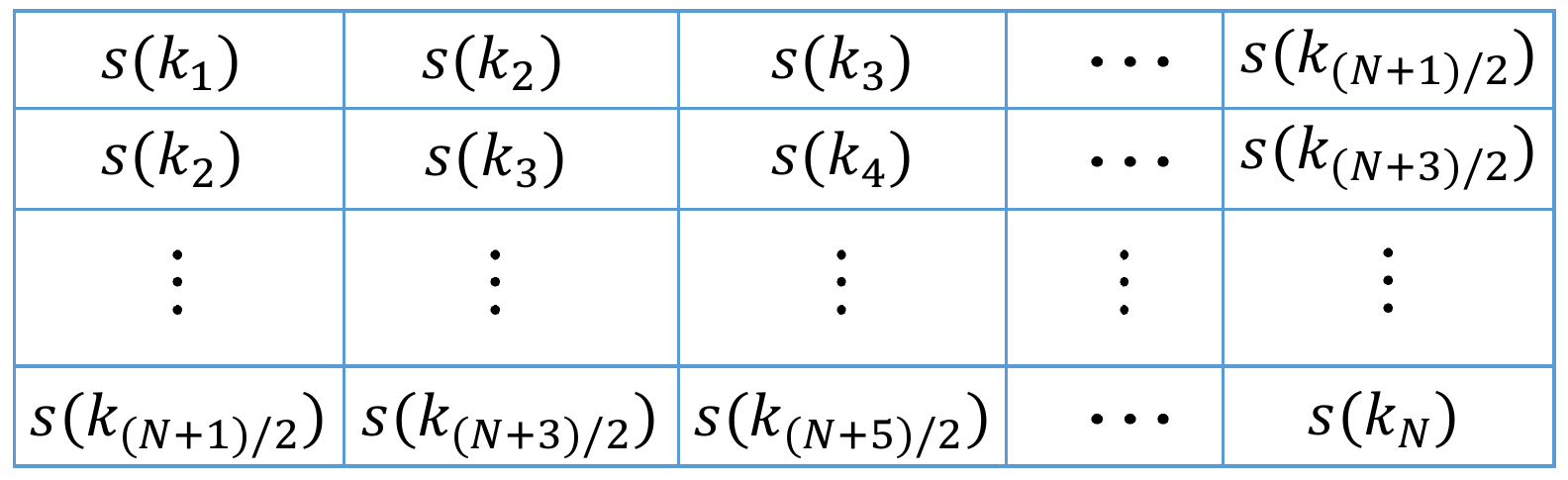}
		\caption{Dimension increasing from $s(k,\cdots)$ to $s(k_T,k_R,\cdots)$.}
		\label{dim_increa}
	\end{figure}

	In order to achieve independent data with respect to $k_{\rho_T}$ and $k_{\rho_R}$, we perform dimension increase by dividing $k$ into two parts using the relation $k=k_T+k_R$, with
	\begin{subequations} \label{dispersion2}
		\begin{align} 
			&k_{\rho_T}=\sqrt{4k^2_{T}-k_{z'}^2/4},\label{kt} \\ 
			&k_{\rho_R}=\sqrt{4k^2_{R}-k_{z'}^2/4}.\label{kr}
		\end{align}
	\end{subequations}
	To achieve this goal, we reformulate the data $s(k,\cdots)$  according to the illustration in Fig. \ref{dim_increa}.
	In so doing, we get a data matrix corresponding to the independent polar spatial frequency support $(k_T,\theta_T)$ and $(k_R,\theta_R)$, respectively.
	
	Then, substituting \eqref{cylindrical2planeT} and \eqref{cylindrical2planeR} in \eqref{scat_wave_kz3}, and rearranging the integral sequence, we obtain,
	\begin{align} \label{scat_wave4}
		&s(k_T,k_R,\theta_T,\theta_R,k_{z'})\!=\!\iint\!\iiint g(x,y,z)\cdot\\ \nonumber
		&e^{-\mathrm{j}(k_{x_T}+k_{x_R})x}e^{-\mathrm{j}(k_{y_T}+k_{y_R})y}e^{-\mathrm{j}k_{z'}z}\cdot\\ \nonumber
		&e^{\mathrm{j}k_{x_T}x_T}e^{\mathrm{j}k_{y_T}y_T}e^{\mathrm{j}k_{x_R}x_R}e^{\mathrm{j}k_{y_R}y_R}\mathrm{d}x\mathrm{d}y\mathrm{d}z\mathrm{d}k_{x_T}\mathrm{d}k_{x_R}. 
	\end{align}
	Clearly, the inner integrals over $x$, $y$, and $z$ can be expressed as a 3-D Fourier transform of $g(x,y,z)$ (denoted by $G(k_x,k_y,k_z)$). Then, we have
	\begin{align} \label{scat_wave5}
		{s}(k_T,k_R,&\theta_T,\theta_R,k_{z})\!=\!\iint G(k_x,k_y,k_z)\cdot\\ \nonumber
		&e^{\mathrm{j}k_{x_T}x_T}e^{\mathrm{j}k_{y_T}y_T}e^{\mathrm{j}k_{x_R}x_R}e^{\mathrm{j}k_{y_R}y_R}\mathrm{d}k_{x_T}\mathrm{d}k_{x_R}, 
	\end{align}
	where 
	\begin{subequations}
		\begin{align} 
			&k_x=k_{x_T}+k_{x_R}, \label{kx} \\ 
			&k_y=k_{y_T}+k_{y_R}, \label{ky}  \\
			&k_z=k_{z'}.
		\end{align}
	\end{subequations}
	
	The exponential terms associated with the integrals in the right side of \eqref{scat_wave5} can be canceled in the cylindrical coordinates. The spatial frequency relations between the Cartesian and the cylindrical coordinates are given by,
	\begin{subequations}
		\begin{align} 
			&k_{x_{T/R}}=k_{\rho_{T/R}}\sin\phi_{T/R}, \label{kxtr}\\
			&k_{y_{T/R}}=-k_{\rho_{T/R}}\cos\phi_{T/R}. \label{kytr} 
		\end{align}
	\end{subequations}
	The differential elements in \eqref{scat_wave5} can then be approximated by  $\mathrm{d}k_{x_T}\mathrm{d}k_{x_R}\approx k_{\rho_T}k_{\rho_R}\cos\phi_T\cos\phi_R\mathrm{d}\phi_T\mathrm{d}\phi_R$ due to the fact that  $k_{\rho_T}$ and $k_{\rho_R}$ vary slowly along the direction of $k_{x_T}$ and $k_{x_R}$, respectively, for a limited angle extent subtended by the circular-arc array. 
	Then, representing the spatial frequencies in the cylindrical coordinates, \eqref{scat_wave5} is rewritten as
	\begin{align} \label{scat_wave7}
		{s}(k_T,k_R,&\theta_T,\theta_R,k_{z_T},k_{z_R})\!=\!k_{\rho_T}k_{\rho_R}\cdot \\ \nonumber
		\!\iint \! &G(k_{\rho_T},k_{\rho_R},\phi_T,\phi_R,k_z)\cos \phi_T\cos \phi_R\cdot\\ \nonumber
		&e^{\mathrm{j}k_{\rho_T}R_0\cos(\theta_T-\phi_T)}e^{\mathrm{j}k_{\rho_R}R_0\cos(\theta_R-\phi_R)}\mathrm{d}\phi_{T}\mathrm{d}\phi_{R}. 
	\end{align}
	Clearly, the integrals over $\phi_{T}$ and $\phi_{R}$ can be represented by the following convolutions with respect to $\theta_T$ and $\theta_R$, respectively.
	\begin{align} \label{scat_wave8}
		{s}(&k_T,k_R,\theta_T,\theta_R,k_{z_T},k_{z_R})\\ \nonumber
		=&k_{\rho_T}k_{\rho_R}G(k_{\rho_T},k_{\rho_R},\theta_T,\theta_R,k_z)\cos\theta_T\cos\theta_R\circledast_T \\ \nonumber
		&e^{\mathrm{j}k_{\rho_T}R_0\cos\theta_T}\circledast_R e^{\mathrm{j}k_{\rho_R}R_0\cos\theta_R}, 
	\end{align}
	where $\circledast_T$ and $\circledast_R$ denote  convolutions in the  $\theta_T$ and $\theta_R$ domains, respectively.
	
	Note from \eqref{scat_wave8} the scattering coefficients of target in the spatial frequency domain can be obtained through deconvolutions of  the two exponential functions. 
	We take the Fourier transforms of both sides of \eqref{scat_wave8} with respect to $\theta_T$ and $\theta_R$, respectively. With the help of the Fourier property of convolution, we can write
	\begin{align} \label{scat_wave9}
		&{{s}}(k_T,k_R,\xi_T,\xi_R,k_{z_T},k_{z_R})\!=\!k_{\rho_T}k_{\rho_R} \widetilde{G}(k_{\rho_T},k_{\rho_R},\xi_T,\xi_R,k_z)\cdot \\ \nonumber
		&H^{(1)}_{\xi_T}(k_{\rho_T}R_0)e^{\mathrm{j}\pi\xi_T/2}H^{(1)}_{\xi_R}(k_{\rho_R} R_0)e^{\mathrm{j}\pi\xi_R/2}, 
	\end{align}
	where $\xi_{T/R}$ denotes the Fourier domain for $\theta_{T/R}$, $\widetilde{G}(\cdots)$ represents the Fourier transform of ${G(\cdots)}\cos\theta_T\cos\theta_R$ with respect to $\theta_T$ and $\theta_R$,  
	and $H^{(1)}_{\xi_{T/R}}$ is the Hankel function of the first kind, $\xi_{T/R}$ order \cite{soumekh}. 
	\begin{equation}
		H^{(1)}_{\xi_{T/R}}(k_{\rho_{T/R}} R_0)=\mathcal{F}_{\theta_{T/R}}[e^{\mathrm{j}k_{\rho_{T/R}}R_0\cos\theta_{T/R}}]e^{-\mathrm{j}\pi\xi_{T/R}/2}.
	\end{equation}
	
	Thus, dividing both sides of \eqref{scat_wave9} by the two Hankel functions and exponentials, and performing the inverse Fouirer transforms with respect to $\xi_{T}$ and $\xi_{R}$, we obtain
	\begin{align} \label{freq_target}
		&G(k_{\rho_T},k_{\rho_R},\theta_T,\theta_R,k_{z})=\frac{1}{k_{\rho_T}k_{\rho_R}\cos\theta_T\!\cos\theta_R}\cdot \\ \nonumber
		&\mathcal{F}^{-1}_{\xi_{T\!/\!R}}\!\!\Bigg[{\frac{{{s}}(k_T,k_R,\xi_T,\xi_R,k_{z_T},k_{z_R})e^{-\mathrm{j}\pi\xi_T/2}e^{-\mathrm{j}\pi\xi_R/2}}{H^{(1)}_{\xi_T}(k_{\rho_T}R_0)H^{(1)}_{\xi_R}(k_{\rho_R} R_0)}}\Bigg].
	\end{align}
	
	Finally, interpolations based on  \eqref{kxtr} and \eqref{kytr} and dimension reduction \cite{zhuge2} are implemented to change $G(k_{\rho_T},k_{\rho_R},\theta_T,\theta_R,k_z)$ into  $G(k_x,k_y,k_z)$, over which the 3-D inverse FFT is performed to obtain $g(x,y,z)$. 
	
	\subsection{Sampling criteria}
	
	\begin{figure}[!t]
		\centering
		
		\includegraphics[width=1.6in,height=1.4in]{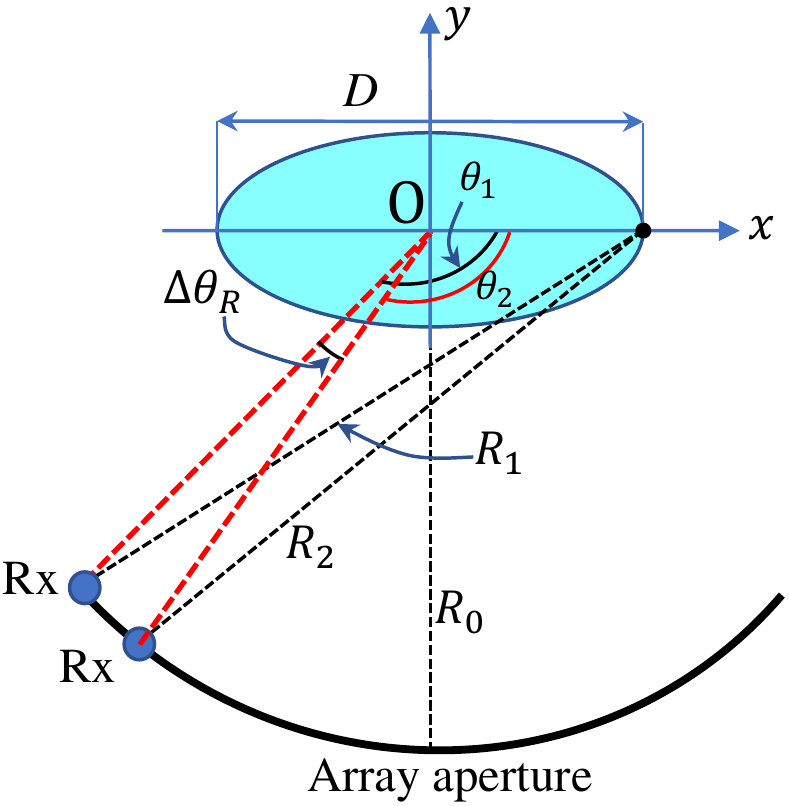}
		
		\caption{Illustration of the sampling criterion of circular-arc MIMO array.}
		\label{sampling_circular}
	\end{figure}
	
	We assume that the transmit array is undersampled, and the receive array is fully sampled. 
	First, the requirement for the inter-element spacing of the receive array is considered. To avoid aliasing, the phase difference between two neighboring
	receive antennas has to be less than $\pi$ rad \cite{zhuge2}. Considering the illustration in Fig. \ref{sampling_circular}, we have $k(R_1-R_2)\leq \pi$, where $R_1=\sqrt{R_0^2+\frac{D^2}{4}-R_0 D\cos\theta_1}\approx R_0-\frac{D\cos \theta_1}{2}$, $R_2=\sqrt{R_0^2+\frac{D^2}{4}-R_0 D\cos\theta_2}\approx R_0-\frac{D\cos \theta_2}{2}$, and $\theta_1=\theta_2+\Delta \theta_R$.
	Thus, 
	\begin{align} \nonumber
		&k(R_1-R_2)\\\nonumber
		&\approx k\frac{D\sin\theta_2\sin\Delta\theta_R}{2} \leq k\frac{D\sin\Delta\theta_R}{2}\approx k\frac{D\Delta\theta_R}{2}\leq\pi, 
	\end{align}
	due to $\cos\Delta\theta_R \approx 1$ and $\sin\Delta\theta_R \approx \Delta\theta_R$ with a small inter-element spacing $\Delta\theta_R$ of the receive array. Then, we obtain
	\begin{equation} \label{spacing_receive}
		\Delta\theta_R\leq \frac{\lambda_{\text{min}}}{D},
	\end{equation}
	where $\lambda_{\text{min}}$ denotes the minimum wavelength of the working EM waves, and $D$ is the maximum dimension along $x$ direction. 
	
\begin{figure}[!t]
		\centering
		
		\includegraphics[width=2.7in]{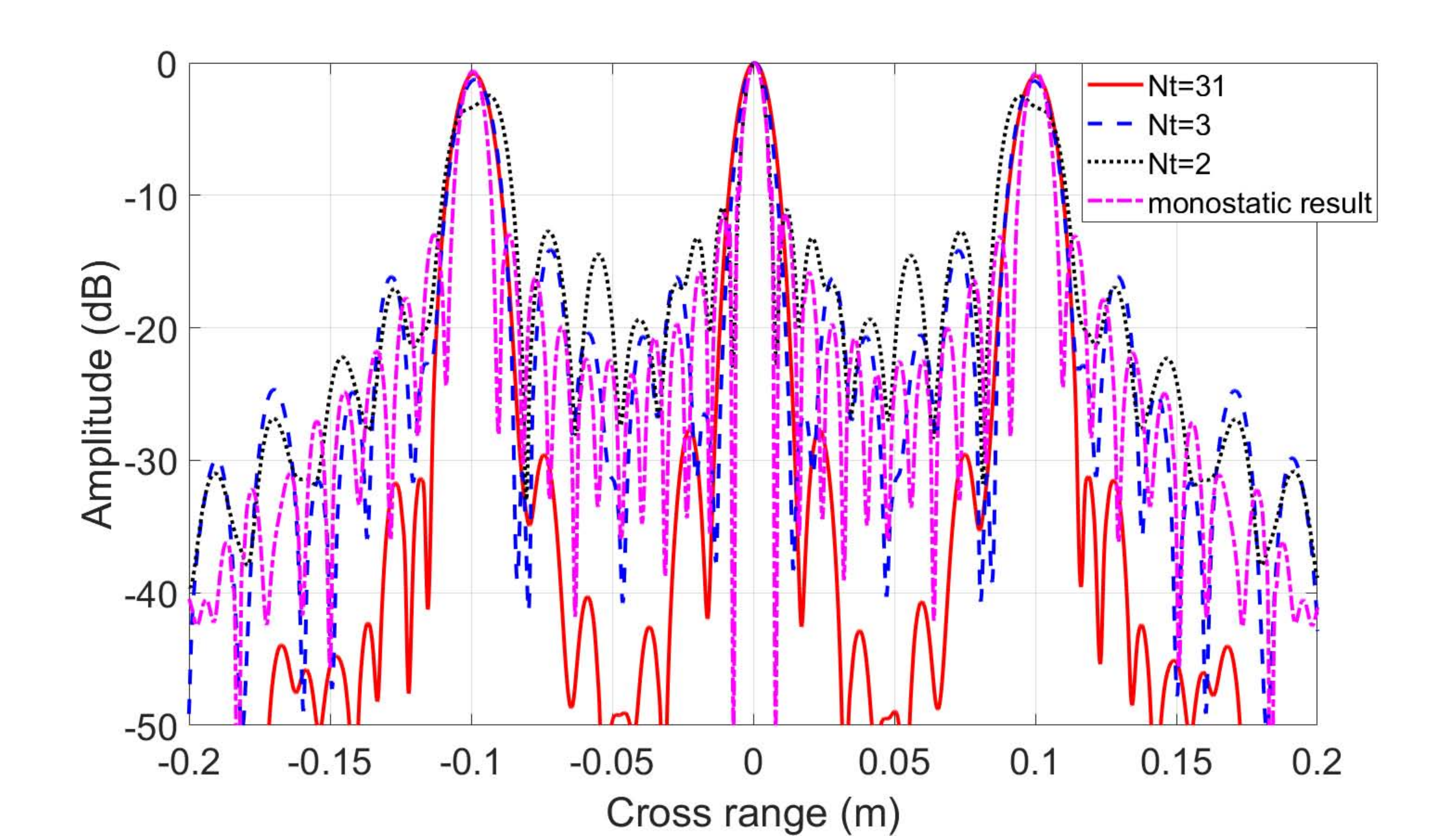}
		
		\caption{1-D images along the horizontal direction by using circular-arc MIMO arrays with $N_t$ transmit antennas and 31 receive antennas, and by using a monostatic array with 61 transmit-receive pairs.}
		\label{comp_Txs}
	\end{figure}

	For the undersampled transmit array, there is no limit to the inter-element spacing as long as two antennas at least are put at the both ends of the receive array to ensure a comparable resolution with a monostatic array. 
	Fig. \ref{comp_Txs} shows a comparison of 1-D imaging results obtained using a monostatic array with 61 transmit and 61 receive antennas as a benchmark, and using circular MIMO arrays with $N_t$ transmit antennas and 31 receive antennas. 
	Note that the resolutions of MIMO arrays are slightly worse than that of the monostatic array due to the convolution in \eqref{scat_wave8}, which is equivalent to adding a weighting function (shaped like  a triangle) to the spatial frequency data. 
	On the other hand, the weighting function with respect to $N_t=2$ is shaped somewhat like a `U'-shape, which results in a slightly narrower main lobe but higher sidelobes than those of the other cases (such as $N_t=3$ or $31$). 
	However, on the whole, the results in Fig. \ref{comp_Txs} have comparable resolutions. 
	

	As for the  requirement for the mechanical scanning step along the vertical direction, according to  $e^{-\text{j}k_z z'}$ in \eqref{scat_wave4}, the scanning step should satisfy $k_{z_{\text{max}}}\Delta z' \leq \pi$ to avoid aliasing effects. Based on  \eqref{dispersion}, we have $k_{z_{\text{max}}} = 2k_{\text{max}}\sin\frac{\Theta_z}{2}$, then, 
	\begin{equation}
		\Delta z' \leq \frac{\lambda_{\text{min}}}{4\sin\frac{\Theta_z}{2}},
	\end{equation}
	where $\Theta_z$ denotes the minimum one between the antenna beamwidth and the angle subtended by the scanning length from the center of imaging scene. 
	
	\subsection{Resolutions}
	
	First,  consider the cross-range resolution along the  direction  of the  circular-arc array. Note that the cross-range resolution is determined by the extent of the corresponding spatial frequency, which leads to,
	\begin{equation}
		\delta x = \frac{\pi}{k_{x_{\text{max}}}},
	\end{equation}
	based on the exponential function $\exp(-\mathrm{j}k_{x}x)$, where $k_{x_{\text{max}}}$ denotes the maximum of $k_x$.
	
	According to \eqref{kx}, and the array configuration in Fig. \ref{circular_mimo}, we note that $k_{x_{\text{max}}}=k_{x_{T_{\text{max}}}}+k_{x_{R_{\text{max}}}}$ with $k_{x_{T_{\text{max}}}}=k_{x_{R_{\text{max}}}}\approx k_c\sin(\Theta_h /2)$ with the help of \eqref{kt}, \eqref{kr}, and \eqref{kxtr}, where $k_c$ denotes the center wavenumber of the working waves, and $\Theta_h$ represents the  angle subtended by the array aperture, assuming that the beamwidth of each antenna can fully illuminate the target in the horizontal direction. Thus, 
	\begin{equation}
		\delta x = \frac{\pi}{2k_c\sin \frac{\Theta_h}{2}} = \frac{\lambda_c}{4\sin\frac{\Theta_h}{2}}.
	\end{equation}
	This is the same with that of the monostatic imaging scenario, which has been verified by the results in Fig. \ref{comp_Txs}. 
	
	Similarly, the resolution along the $z$ direction is given by,
	\begin{equation}
		\delta z = \frac{\pi}{k_{z_{\text{max}}}}=\frac{\lambda_c}{4\sin\frac{\Theta_z}{2}},
	\end{equation}
	where $\Theta_z$ denotes the minimal one between the angle subtended by the scanning length and the beamwidth of the antenna element. 
	

	Finally, the down-range resolution can be expressed as $\delta y = \frac{c}{2B}$, where $B$ denotes the bandwidth of the working EM waves.
	
	\section{Results}
	
	\begin{table}
		\centering
		\caption{Simulation Parameters}
		\label{tab3}
		\setlength{\tabcolsep}{3pt}
		\begin{threeparttable}
			\begin{tabular}{p{160pt}  p{30pt}}
				\hline\hline
				Parameters& Values \\[0.5ex]
				\hline
				Radius of the circular array $(R_0)$&
				1.0 m\\[0.5ex]
				Start frequency& 
				30 GHz \\[0.5ex]
				Stop frequency&
				35 GHz \\[0.5ex]
				Number of frequency steps&
				25 \\[0.5ex]
				Number of transmit antennas&
				5 \\[0.5ex]
				Number of receive antennas&
				41 \\[0.5ex]
				Interval of transmit antennas along circumference&
				9.9 cm \\[0.5ex]
				Interval of receive antennas along circumference&
				0.99 cm \\[0.5ex]
				Scanning step along elevation&
				1.0 cm \\[0.5ex]
				
				\hline
			\end{tabular}
			\label{tab2}
		\end{threeparttable}
	\end{table}

	\begin{figure}[!t]
		\centering
		\subfloat[]{\label{a}
			\includegraphics[width=1.68in]{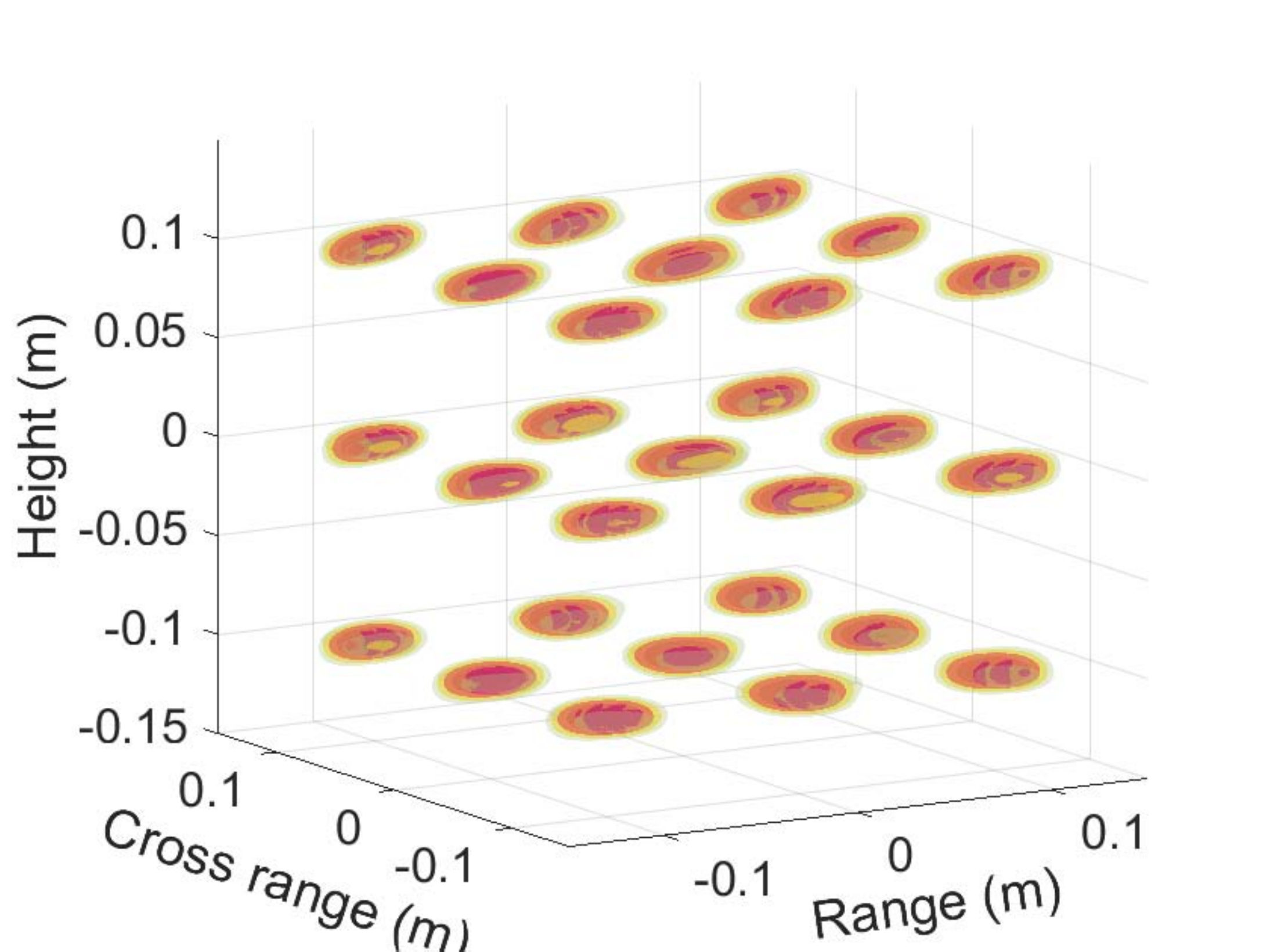}}
		\hfill
		\subfloat[]{\label{b}
			\includegraphics[width=1.68in]{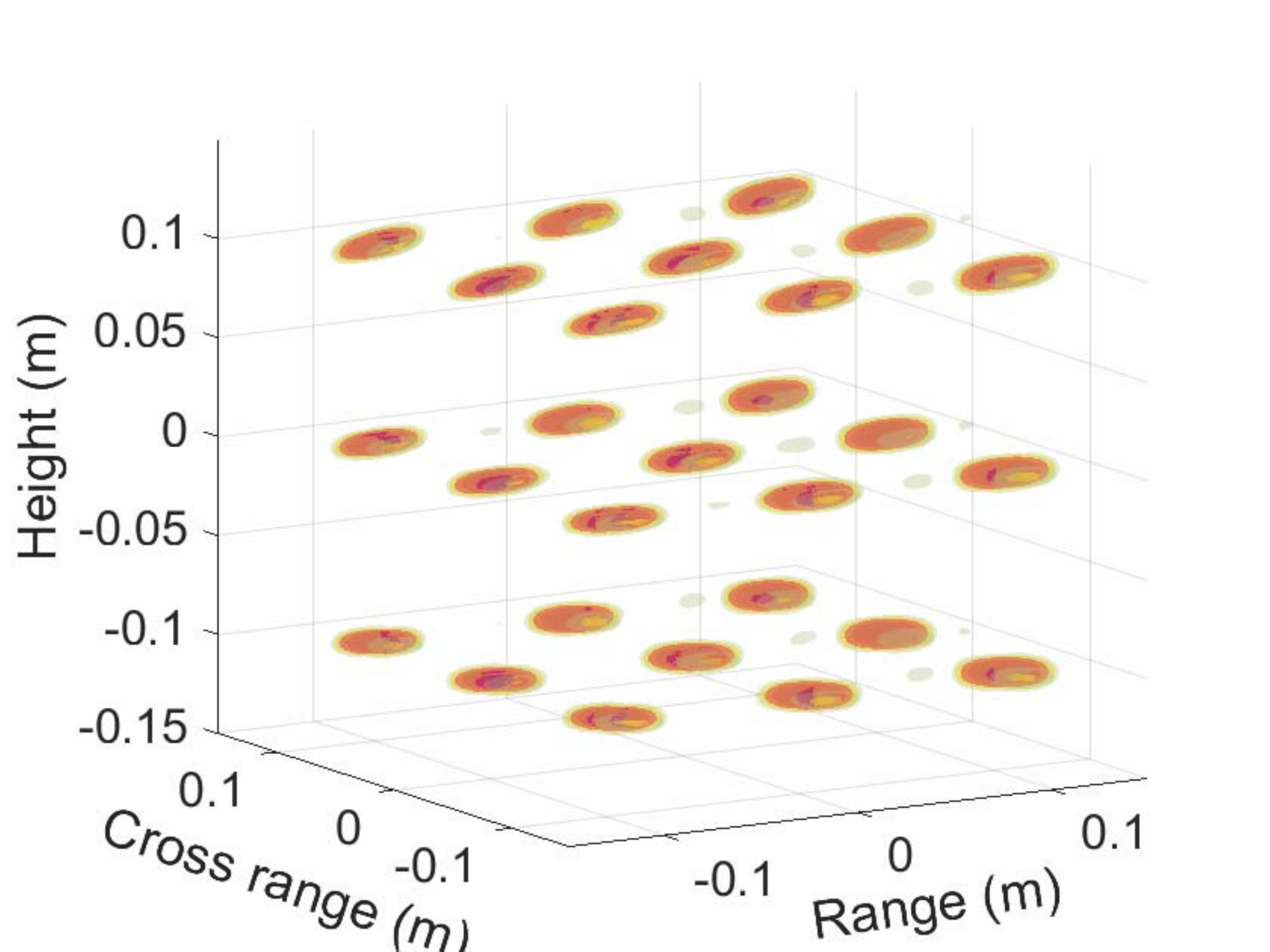}}
		\hfill
		\\	
		\caption{3-D imaging results by (a) the proposed algorithm, and (b) BP, with a dynamic range of 20 dB.}
		\label{3d_mimo}
	\end{figure}

	\begin{figure}[!t]
		\centering
		\subfloat[]{\label{a}
			\includegraphics[width=1.68in]{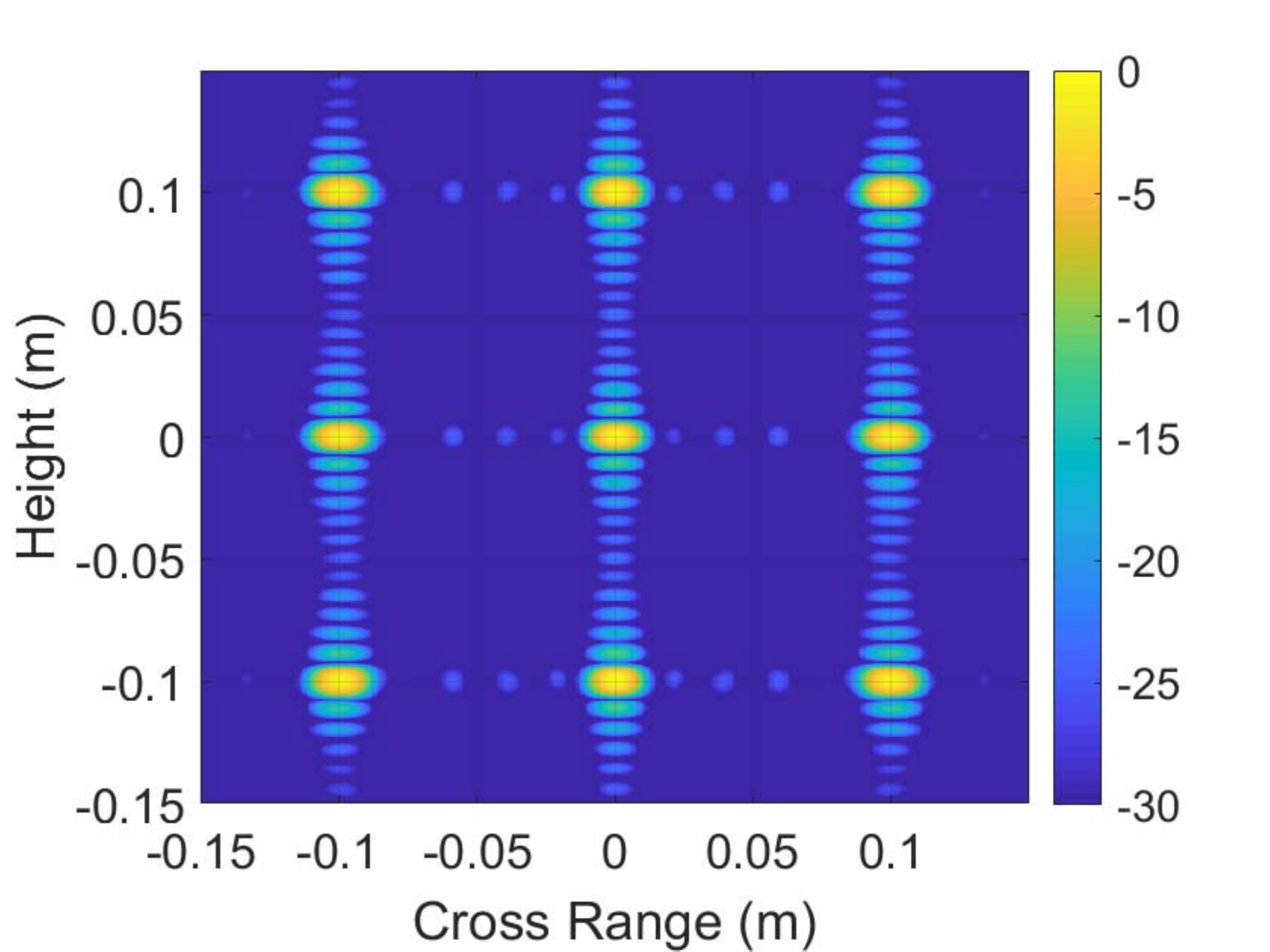}}
		\hfill
		\subfloat[]{\label{b}
			\includegraphics[width=1.68in]{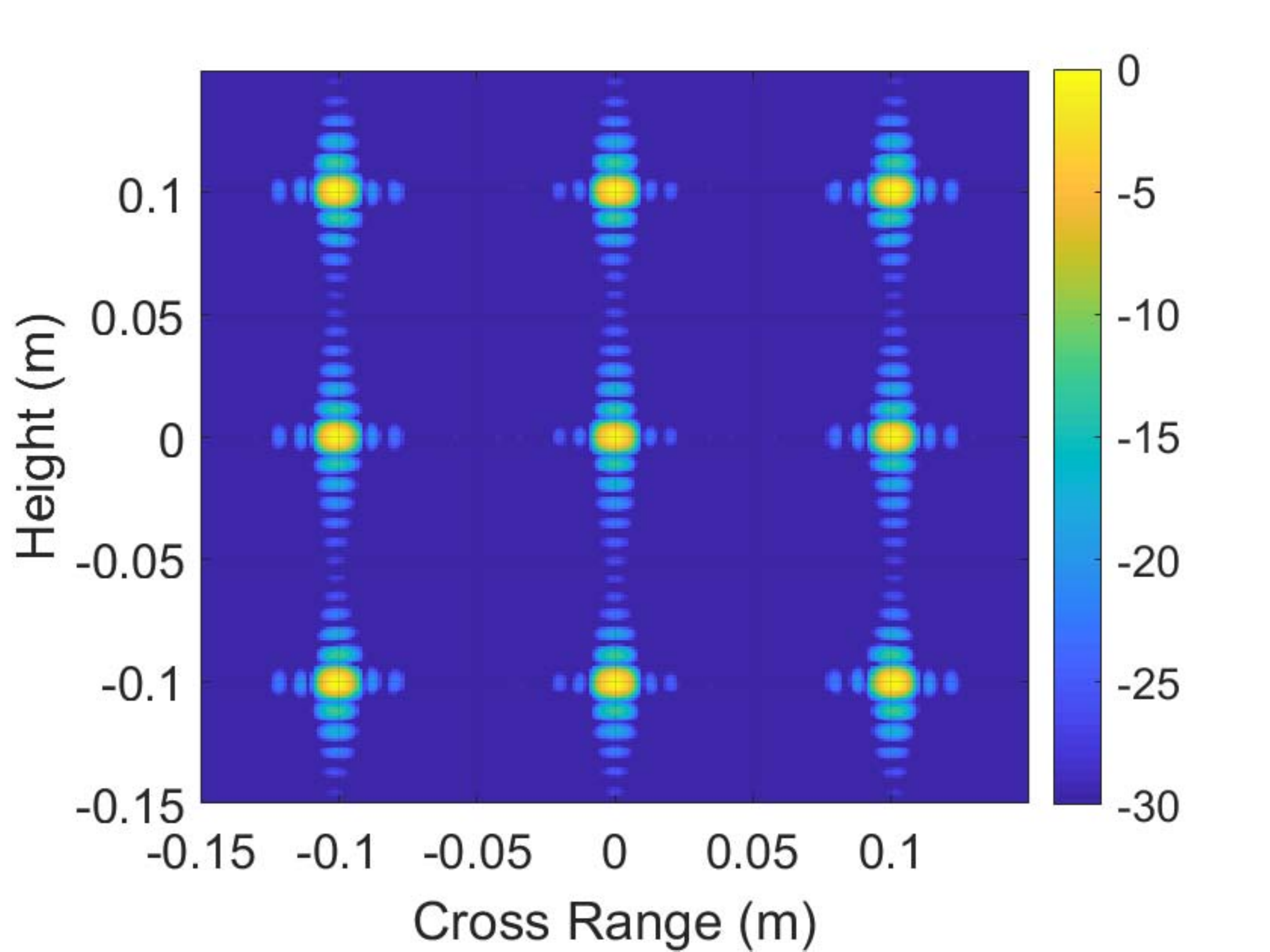}}
		\hfill
		\subfloat[]{\label{c}	
			\includegraphics[width=1.68in]{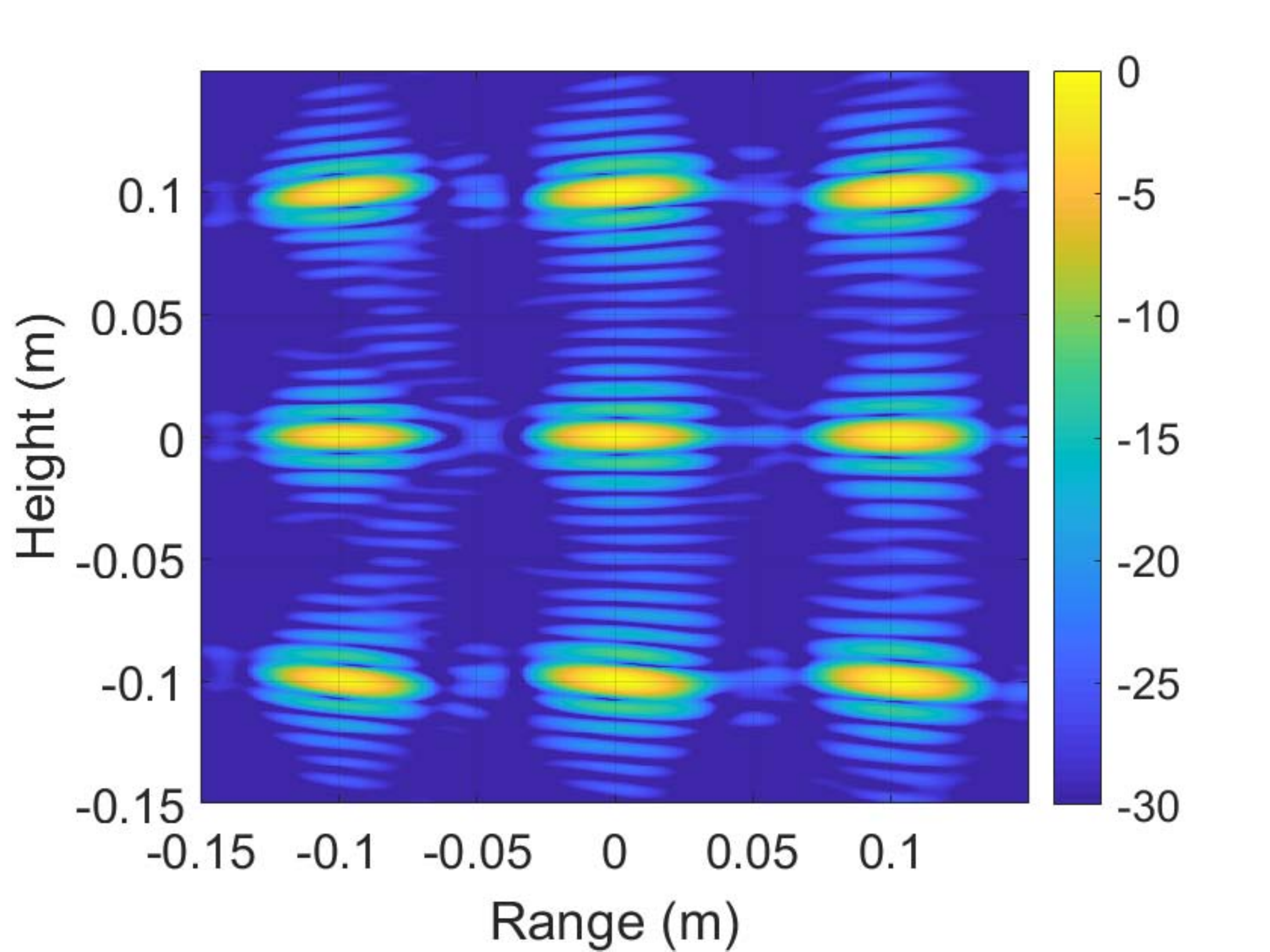}}	
		\subfloat[]{\label{d}
			\includegraphics[width=1.68in]{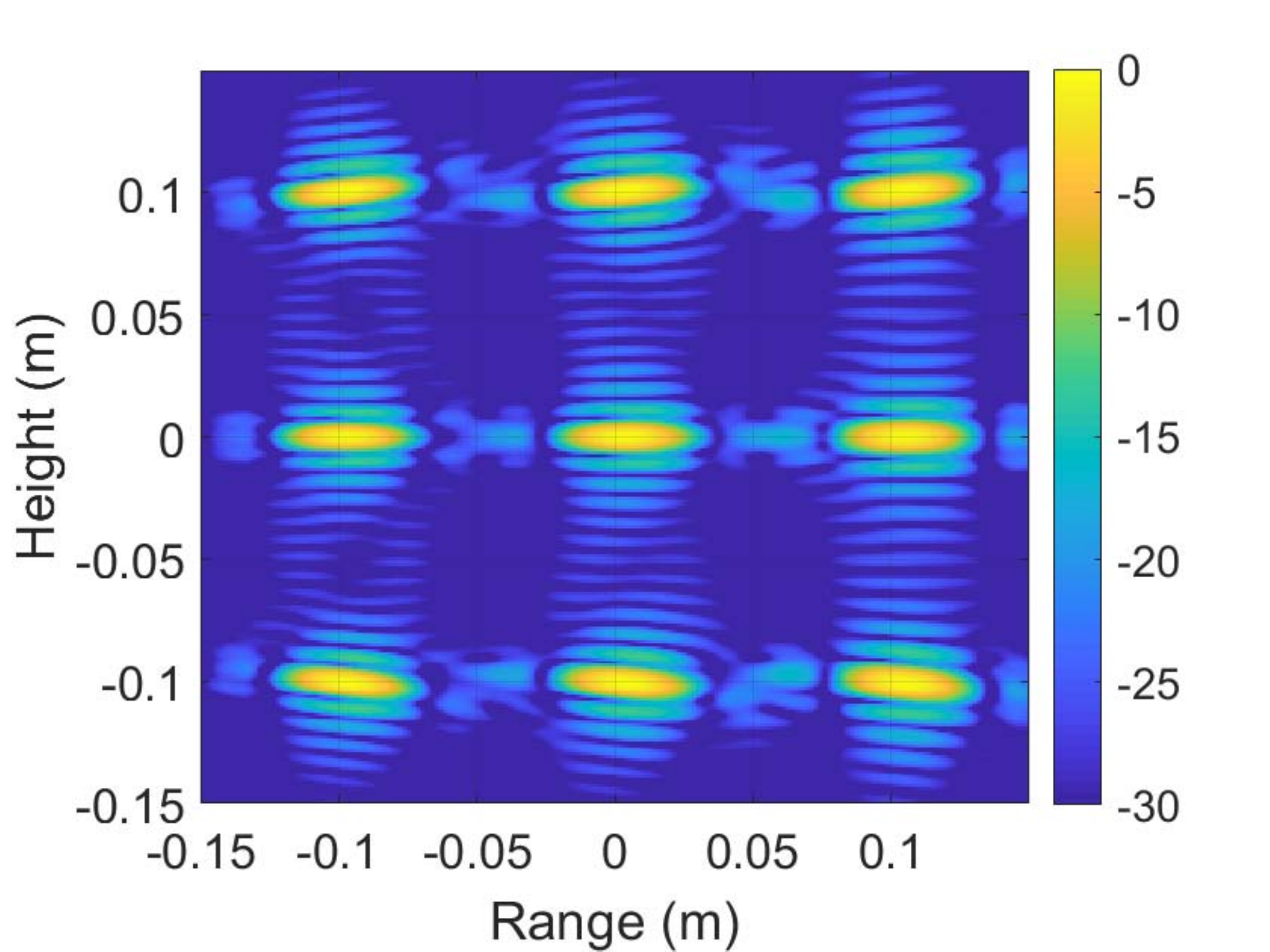}}
		\hfill
		\subfloat[]{\label{e}
			\includegraphics[width=1.68in]{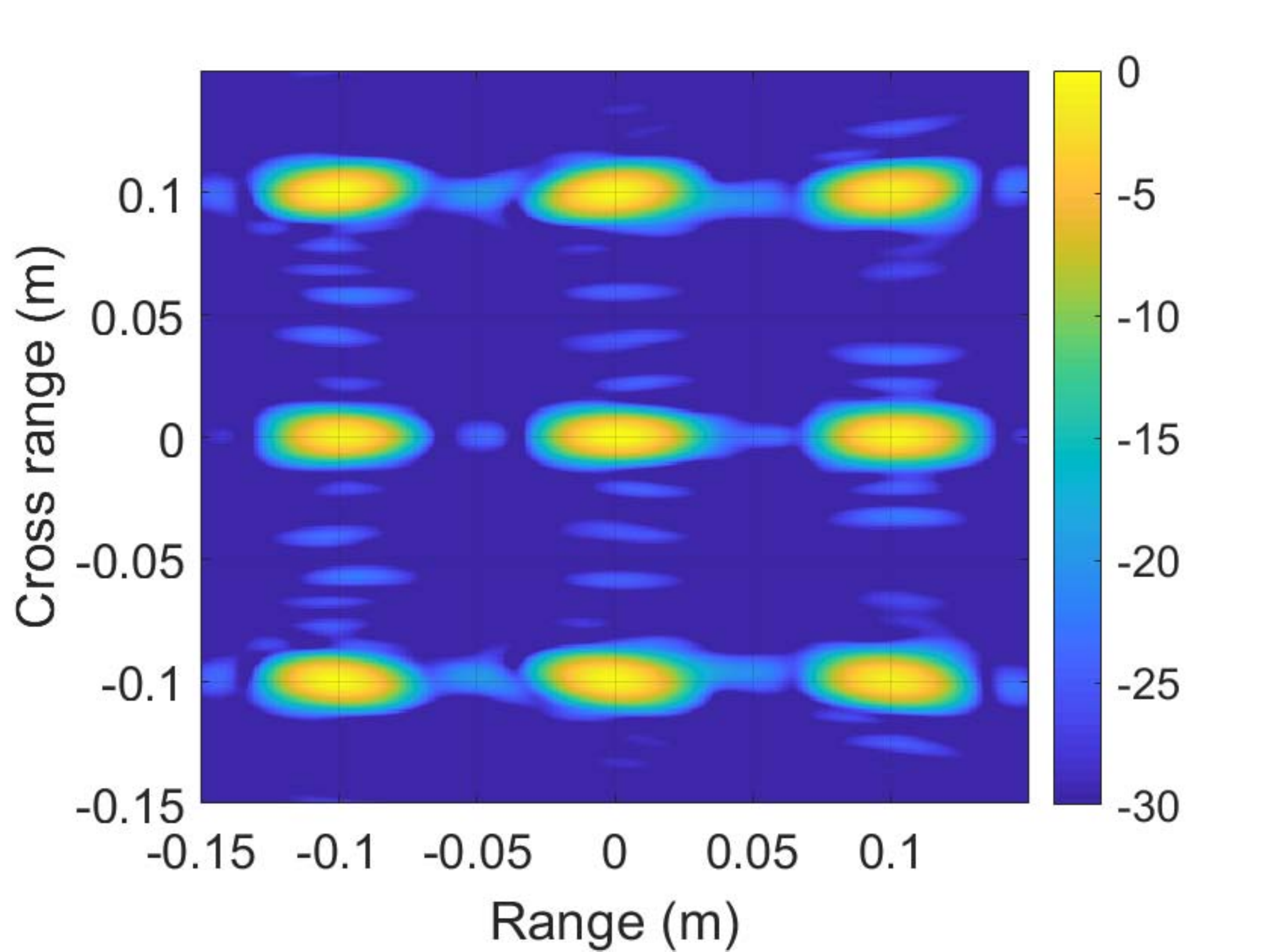}}
		\hfill
		\subfloat[]{\label{f}	
			\includegraphics[width=1.68in]{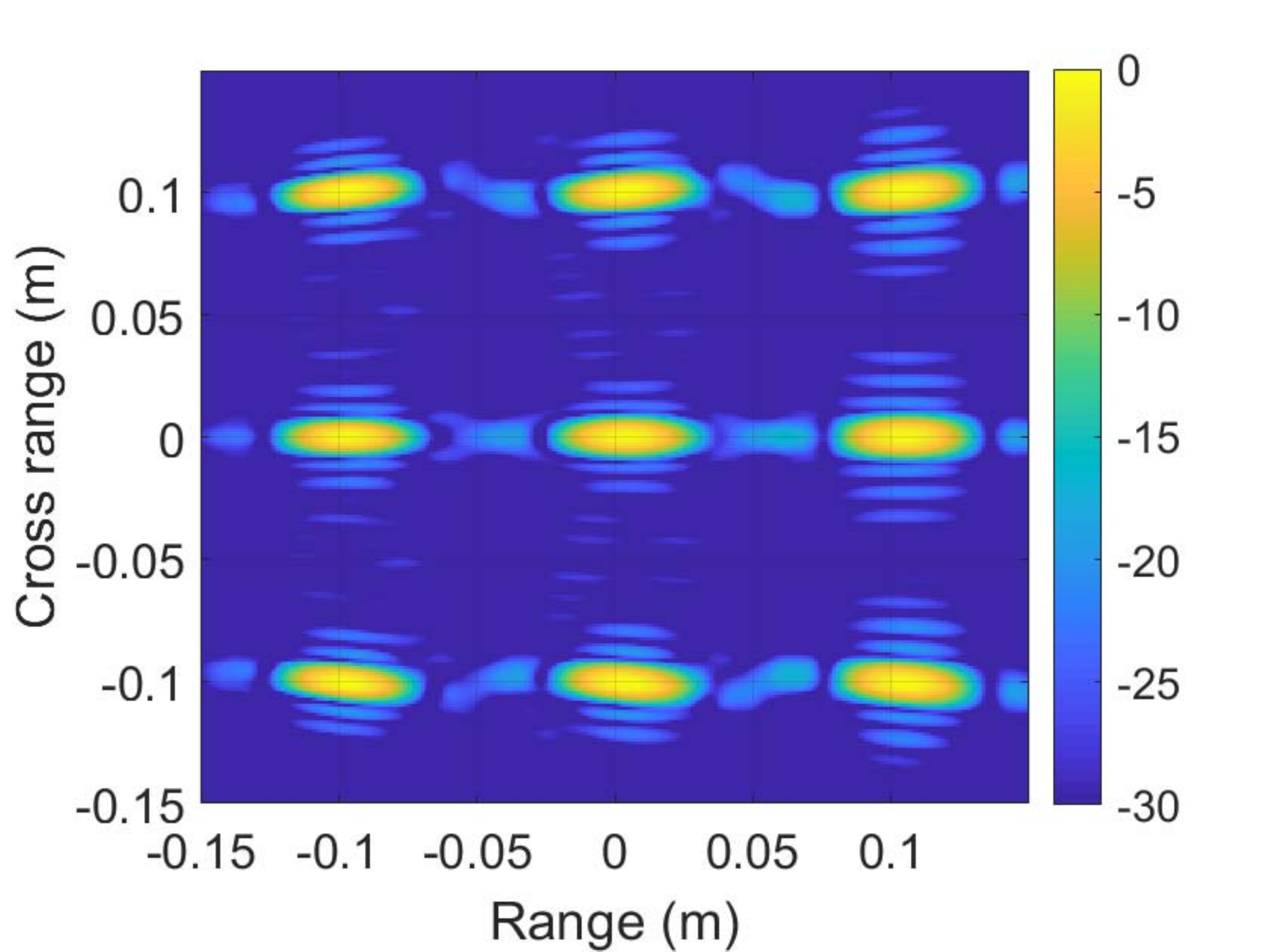}}	
		\\	
		\caption{2-D images  with respect to the three coordinate planes by the proposed algorithm: (a), (c),  (e); and by BP: (b), (d), and (f).  }
		
		\label{2-D slices}
	\end{figure}
	
	\begin{figure}[!t]
		\centering
		\subfloat[]{\label{a}
			\includegraphics[width=2.2in]{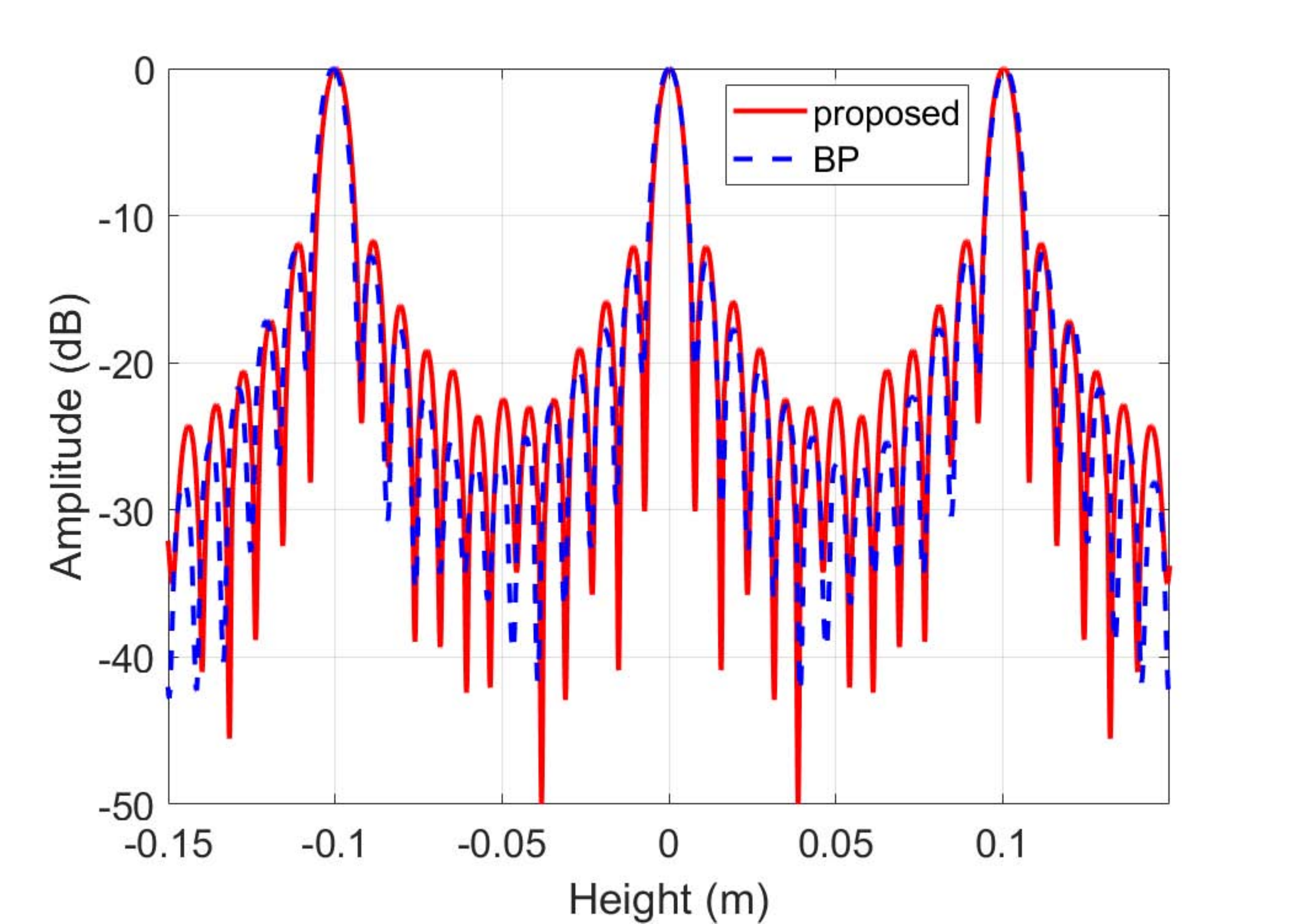}}
		\hfill
		\subfloat[]{\label{b}
			\includegraphics[width=2.2in]{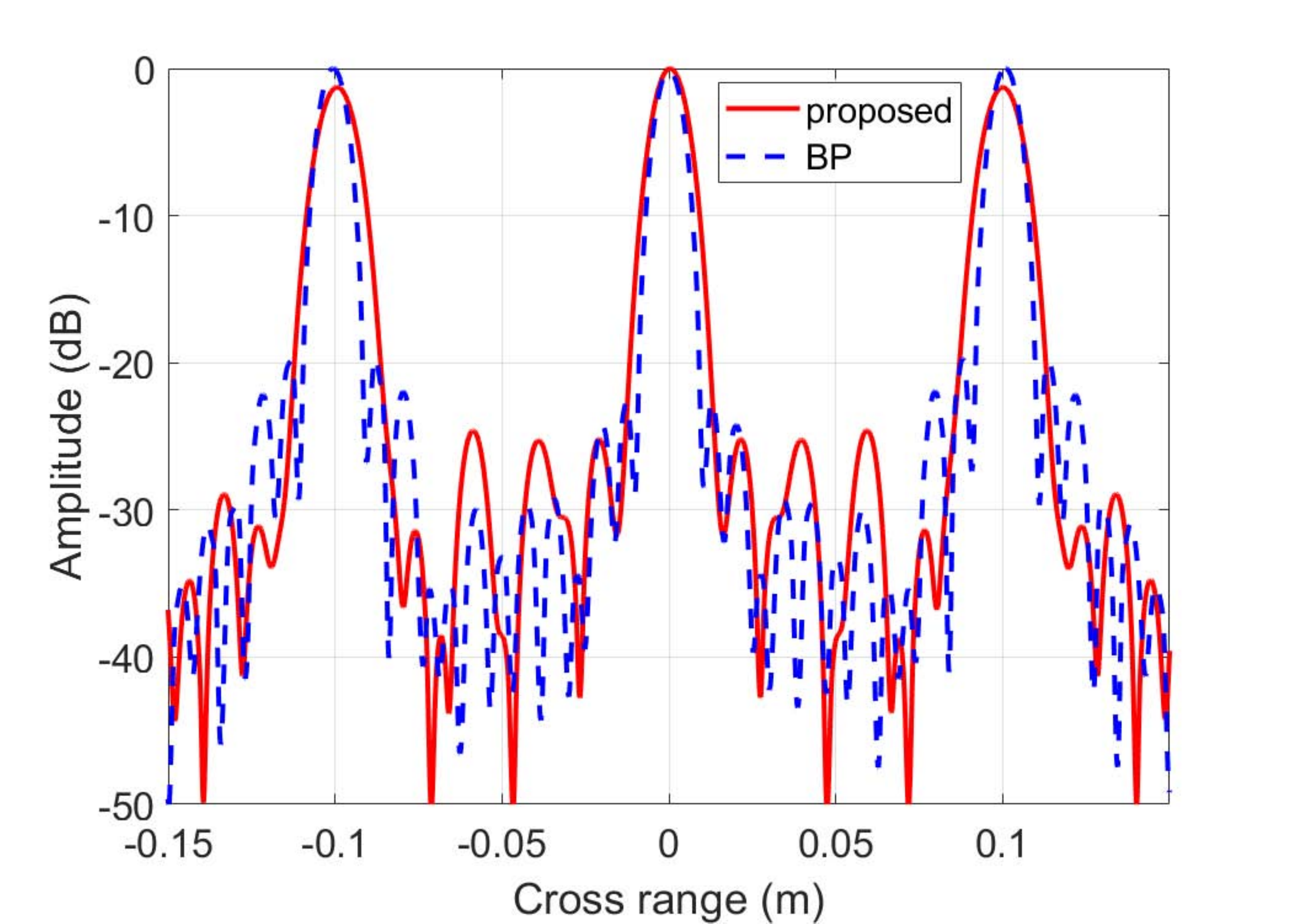}}
		\hfill
		\subfloat[]{\label{c}	
			\includegraphics[width=2.2in]{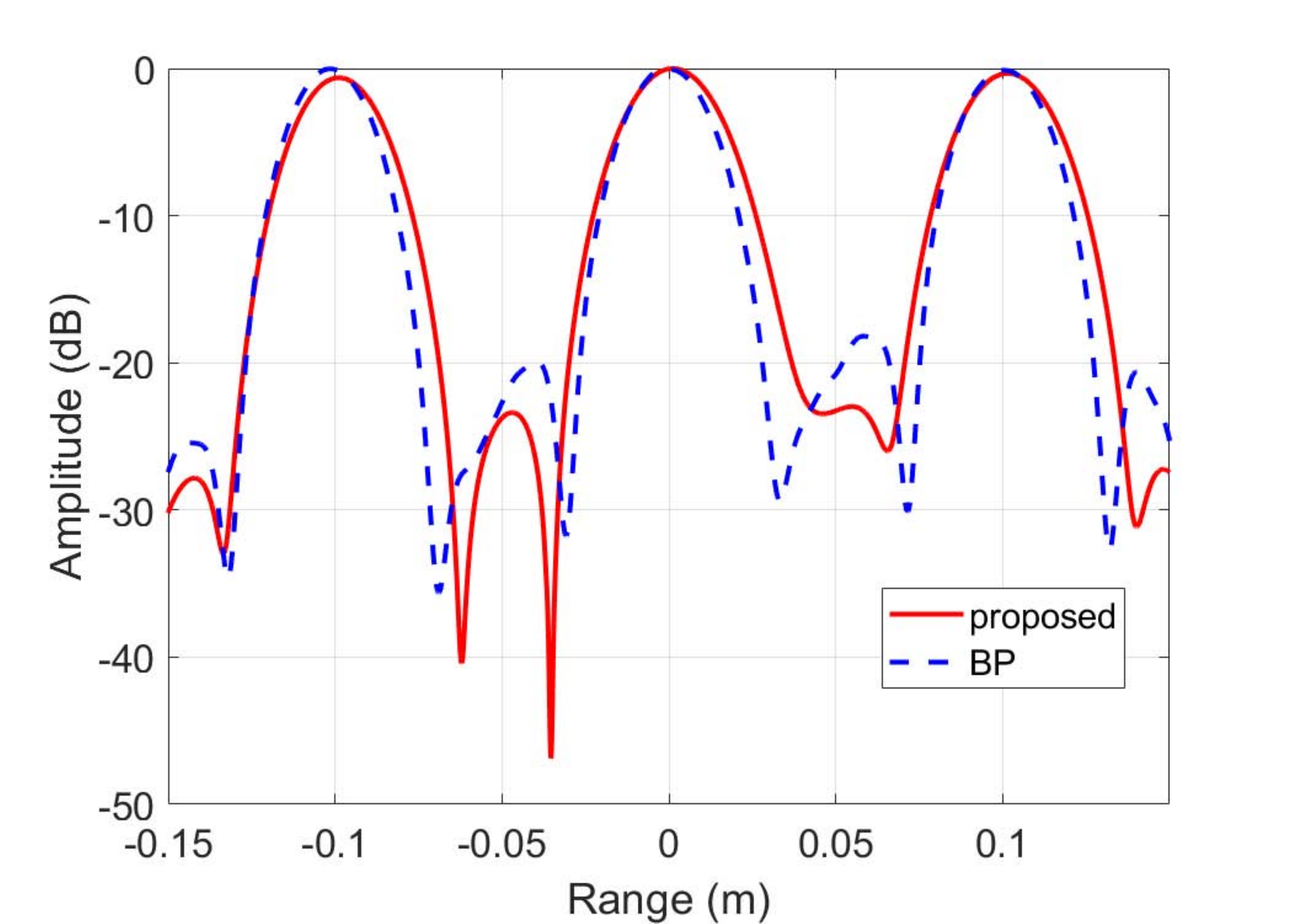}}	
		\\	
		\caption{1-D images corresponding to the (a) height, (b) horizontal, and (c) range dimensions, respectively.}
		\label{1-D slices}
	\end{figure}
	
	In this section, we show more results of the proposed method in comparison with BP.
	The parameters for simulations  are presented in Table \ref{tab2}.
	
	First, we provide simulations of point  targets to compare focusing property with the BP algorithm. 
	The 3-D imaging results are shown in Fig. \ref{3d_mimo}. 
	To view the details, the 2-D images with respect to the three coordinate planes are demonstrated in Fig. \ref{2-D slices}.
	Clearly, the results of the proposed algorithm have similar performance with those of BP. 
	Fig. \ref{1-D slices} shows the 1-D images where the main lobes and sidelobes of the point targets can be clearly demonstrated. 
	Note that the resolutions along the horizontal cross-range and down-range directions in Figs. \ref{1-D slices}(c) and \ref{1-D slices}(d), respectively, of the proposed algorithm are slightly worse than  those of BP, probably due to the information loss caused by the transformation from the polar to Cartesian grids.

\begin{figure}[!t]
		\centering
		\subfloat[]{\label{a}
			\includegraphics[width=1.69in]{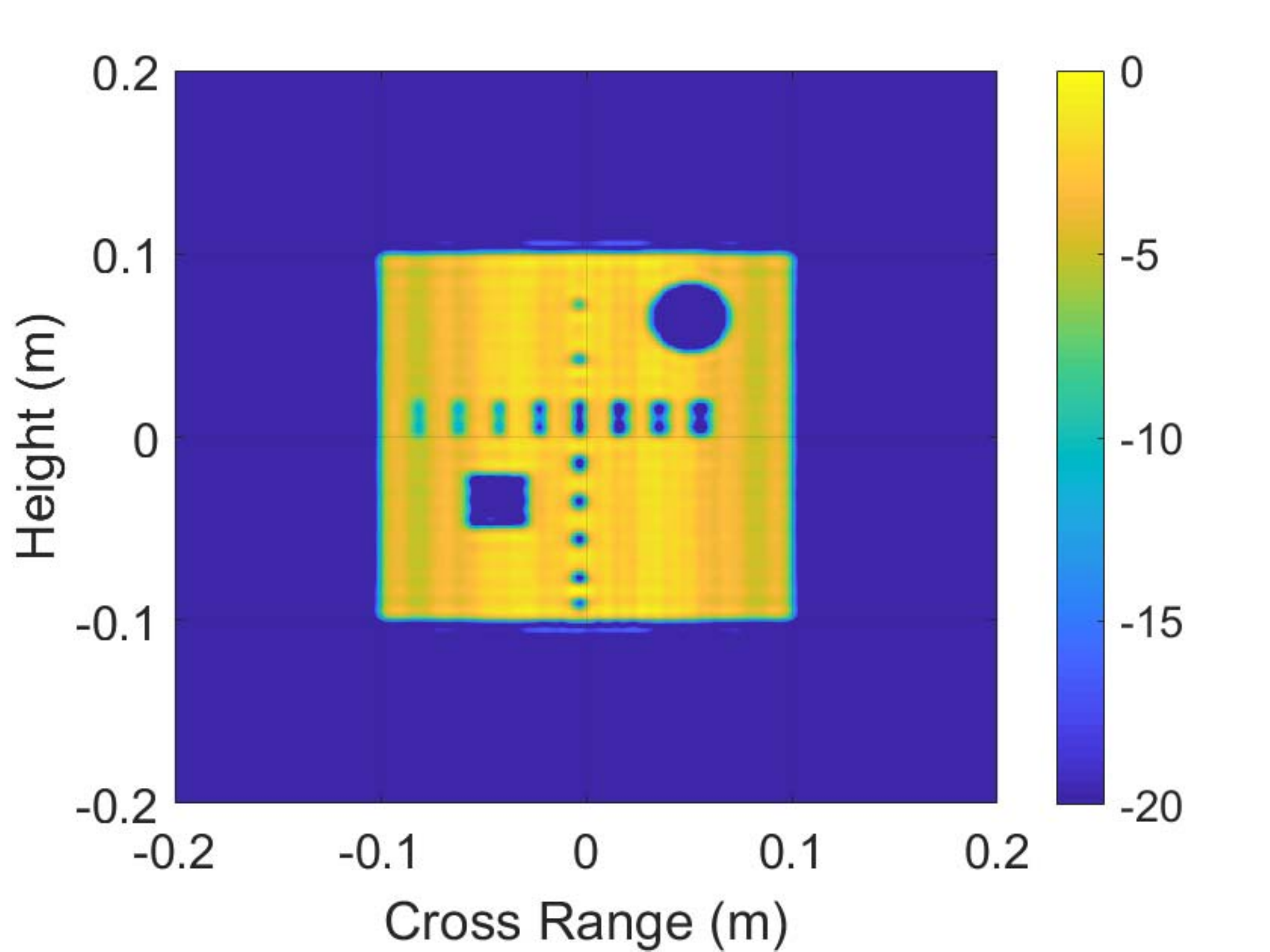}}
		\hfill
		\subfloat[]{\label{b}
			\includegraphics[width=1.69in]{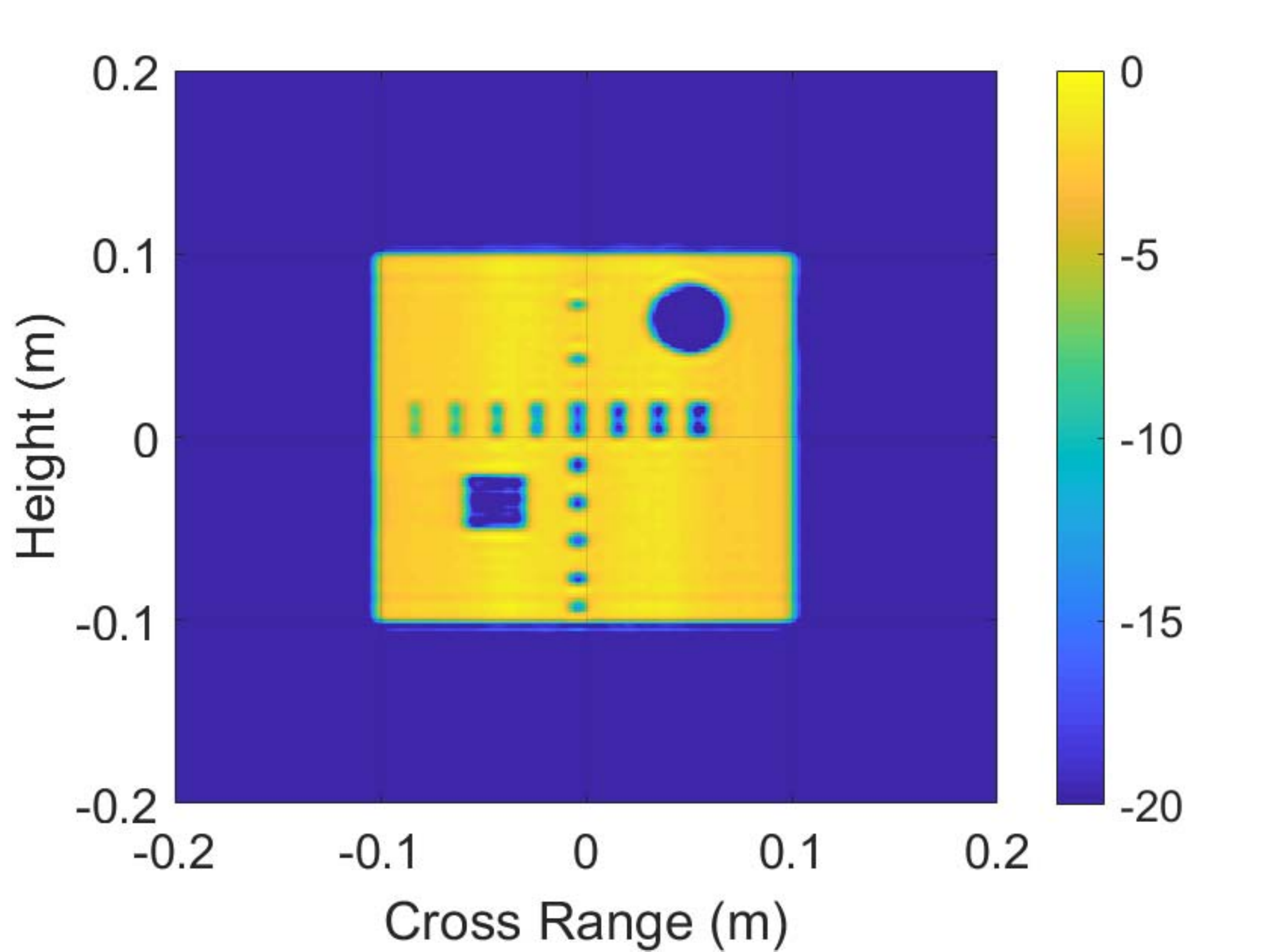}}
		\hfill
		\\	
		\caption{Imaging results by (a) the proposed algorithm, and (b) BP.}
		\label{feko}
	\end{figure}

	Finally, we provide the results by using FEKO - a computational electromagnetics software \cite{feko}, to simulate the scattered EM waves. 
	The reconstructed images by the proposed algorithm and  BP  are shown in Figs. \ref{feko}(a) and \ref{feko}(b), respectively.
	Note that the focusing performance of the proposed algorithm is very close to that of BP, indicating further the effectiveness of the approach.  
	
	\section{Conclusions}
	The paper presented a near-field 3-D MMW imaging scheme based on a circular-arc MIMO array associated with mechanical scanning along its perpendicular direction. The transmit and receive antennas are uniformly distributed over an arc of a circle. The circular-arc MIMO array can provide more even illuminations of the imaging area than the linear or planar MIMO arrays, which may offer better observation of the human
	body and any concealed items. Further, the imaging algorithm was presented based on the 
	spatial frequency domain processing.  
	Numerical experiments demonstrated the efficacy  of the proposed imaging technique.
	
	
	


	\ifCLASSOPTIONcaptionsoff
	\newpage
	\fi

	\bibliography{IEEEabrv,full}
	\bibliographystyle{IEEEtran}
	

\end{document}